\documentstyle[11pt,moriond,epsfig]{article}

\def\be{\begin{equation}}
\def\ee{\end{equation}}
\def\bea{\begin{eqnarray}}
\def\eea{\end{eqnarray}}

\begin{document}
\title{A NEW QUANTUM PHASE IN TWO DIMENSIONS}

\author{ GIULIANO BENENTI, XAVIER WAINTAL, AND JEAN-LOUIS PICHARD }

\address{CEA, Service de Physique de l'Etat Condens\'e,\\
Centre d'Etudes de Saclay, F-91191 Gif-sur-Yvette, France} 

\maketitle\abstracts{
For intermediate Coulomb energy to Fermi energy ratios $r_s$, 
spinless fermions in a random potential form a new quantum phase 
which is different from the Fermi glass and the Wigner crystal. 
From a numerical study of small clusters, we show that this phase 
is characterized by an ordered flow of enhanced persistent currents
and occurs for $r_s$ values where a metallic phase has been observed 
in two dimensions.}

\section{Introduction}
%
%
 
A crucial parameter for a system of charged particles is the Coulomb 
energy to Fermi energy ratio $r_s$. In a disordered two-dimensional 
system, the ground state is obvious in the limiting cases. For large $r_s$, 
the charges form a pinned Wigner crystal, the Coulomb repulsion 
being dominant over the kinetic energy and the disorder. For small 
$r_s$, the interaction becomes negligible and the ground state is a Fermi 
glass with Anderson localized one electron states. 
For intermediate $r_s$, a lot of transport measurements following the 
pioneering works of Kravchenko and co-workers~\cite{kravchenko} and made 
with electron and hole gases give evidence of a metallic phase 
in two dimensions, observed~\cite{hamilton} for instance when 
$6 < r_s < 9$ for a hole gas in GaAs heterostructures. 
In a recent paper~\cite{condmat} we have shown that  
spinless fermions with Coulomb repulsion in disordered $2d$ 
clusters exhibits, for a similar range of $r_s$ values, a new ground state 
characterized by an ordered flow of enhanced persistent currents. 
A study over the statistical ensemble of the currents 
supported by the ground state gives us two well defined values  
$r_s^{F}$ and $r_s^{W}$ bounding this new phase.

\begin{figure} 
\centerline{\epsfxsize=16.0cm\epsfysize=20.0cm\epsffile{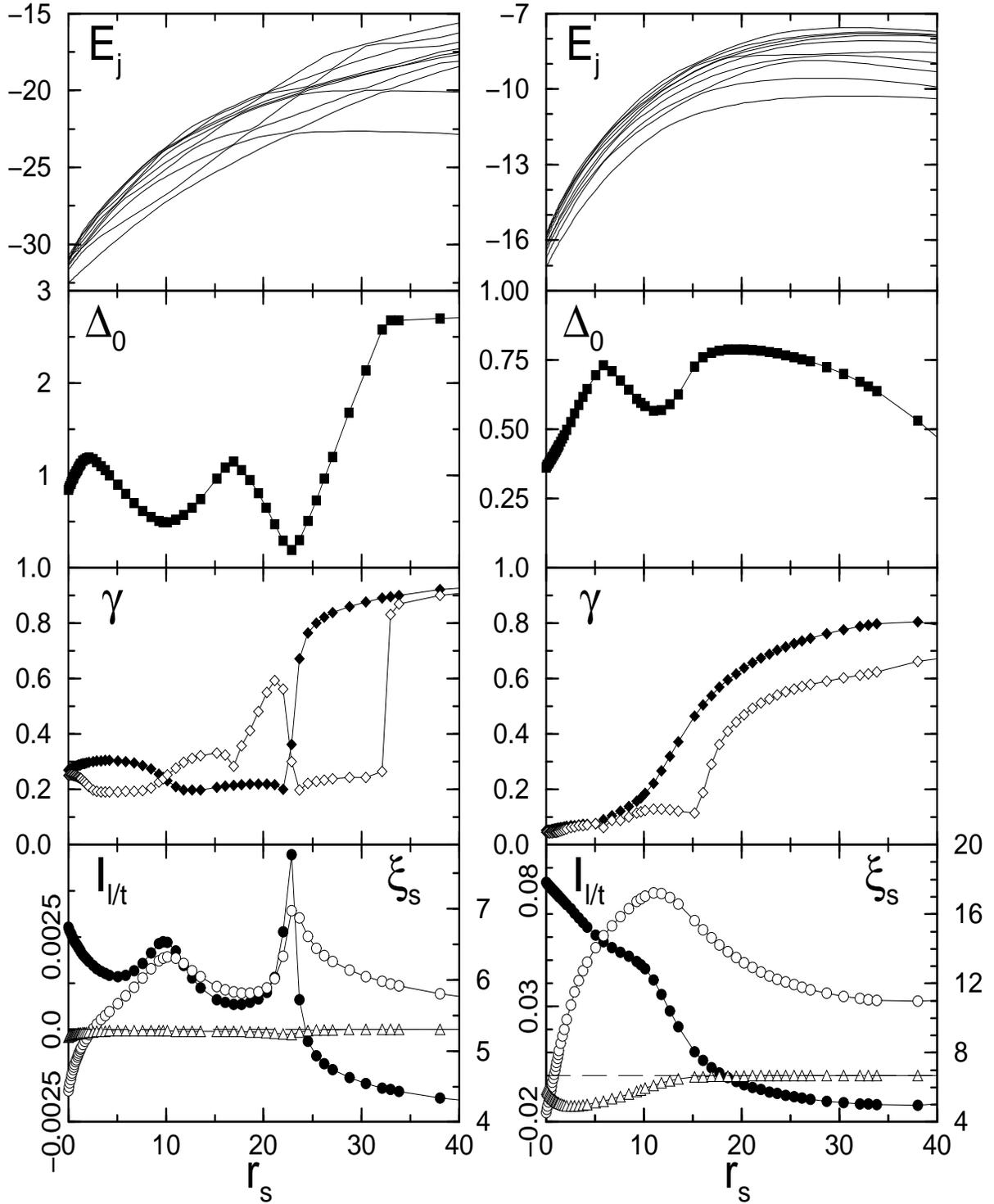}}
\caption{Behavior of a single cluster for $W=15$ (left column) and $W=5$
(right column) as a function of $r_s$. 
From top to bottom: 
low energy spectrum (a $1.9 r_s$ term has been subtracted); 
first energy spacing $\Delta_0$; 
crystallization parameter $\gamma$ for the ground state (filled 
diamonds) and for the first excited state (open diamonds); 
longitudinal $I_l$ (empty circles) and transverse $I_t$ (empty 
triangles) current (left scale) and number of occupied sites $\xi_s$ 
(filled circles, right scale).
\label{fig1}}
\end{figure}

\section{The model} 

We consider a simple model of $N=4$ Coulomb interacting spinless fermions 
in a random potential defined on a square lattice with $L^2=36$ sites. 
The Hamiltonian reads: 
\begin{equation} 
\label{hamiltonian} 
H=-t\sum_{<i,j>} c^{\dagger}_i c_j +  
\sum_i v_i n_i +U \sum_{i\neq j} \frac{n_i n_j}{2 r_{ij}},  
\end{equation} 
where $c^{\dagger}_i$ ($c_i$) creates (destroys) an electron in 
the site $i$, $t$ is the strength of the hopping terms 
between nearest neighbours ($t=1$ in the following) and $r_{ij}$ 
is the inter-particle distance for a $2d$ torus. The random potential 
$v_i$ of the site $i$ with occupation number $n_i=c^{\dagger}_i c_i$ 
is taken from a box distribution of width $W$. The interaction strength 
$U$ yields a Coulomb energy to Fermi energy ratio 
$r_s=U/(2t\sqrt{\pi n_e})$ for a filling factor $n_e=N/L^2$. 
A Fermi golden rule approximation for the elastic scattering time   
leads to $k_F l \approx 192 \pi n_e (t/W)^2$.    
Here $k_F$ and $l$ denote the Fermi wave vector and the elastic mean 
free path respectively and the above estimate is valid  
for filling factors $n_e\ll 1$.   
We consider disorder strengths  
$W=5, 10, 15$ corresponding to $k_Fl= 2.7, 0.7$ and 
$0.3$ respectively. 
The boundary conditions are always taken periodic in the transverse 
$y$-direction, and such that the system encloses an 
Aharonov-Bohm flux $\phi$ in the longitudinal $x$-direction. Imposing 
$\phi=\pi/2$ ($\phi=\pi$ corresponds to anti-periodic boundary 
conditions), one drives a persistent current of total longitudinal 
and transverse components given by 
\begin{equation}
I_{l}=- \frac{\partial E}{\partial \phi} 
=\frac{\sum_i I_i^l}{L}
\,\,\,\,{\rm and}\,\,\,\, 
I_{t}=\frac{\sum_i I_i^t}{L}.  
\end{equation}
The local current $I_i^l$ flowing at the site $i$ in the longitudinal 
direction is defined by $I_i^l =2 {\rm Im} (\langle \Psi_0 | 
c^{\dagger}_{i_{x+1},i_y} c^{}_{i_x,i_y} | \Psi_0 \rangle)$, with an    
analogous expression for $I_i^t$.  
The response is paramagnetic if $I_{l}>0$ and diamagnetic if $I_{l} < 0$. 

%

\begin{figure} 
\centerline{
\epsfxsize=6.5cm\epsfysize=6.5cm\epsffile{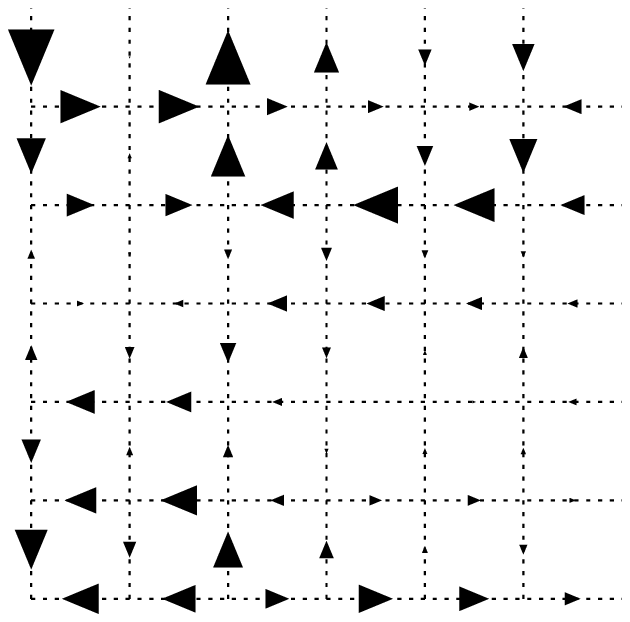}
\hspace{0.8cm}
\epsfxsize=6.5cm\epsfysize=6.5cm\epsffile{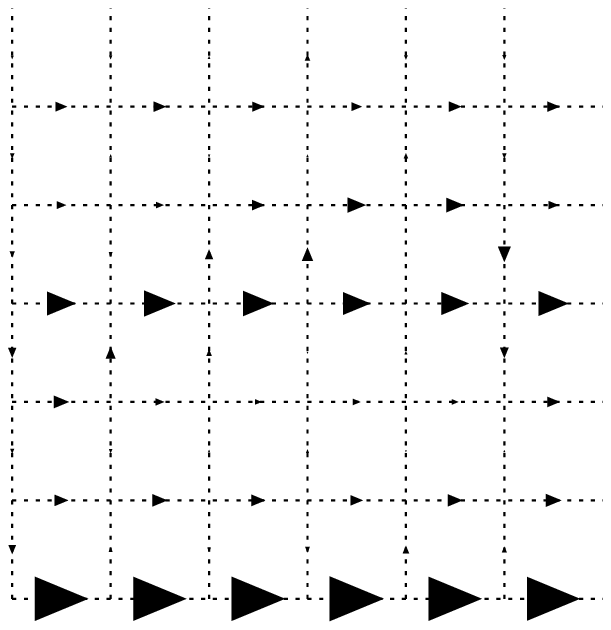}
}
\caption{Local currents for the sample of Fig.~\ref{fig1} at $W=5$,   
$r_s=0$ (left) and $r_s=17$ (right). 
\label{fig2}}
\end{figure}

\section{Numerical results} 

Fig.~\ref{fig1} exhibits behaviours characteristic of individual 
small clusters, comparing the strongly disordered regime $k_F l<1$ 
($W=15$) with the regime $k_F l>1$ ($W=5$). 
For $k_F l<1$, looking at the low energy part of the spectrum 
one can see that, as we gradually turn on the interaction, classification 
of the levels remains invariant up to first avoided crossings, where a 
Landau theory of the Fermi glass is certainly no longer possible. 
The electronic density $\rho_i = \langle \Psi_0| n_i| \Psi_0 \rangle$ of 
the ground state $|\Psi_0\rangle$ is  
maximum in the minima of the site potentials for the Fermi glass. After the 
second avoided crossing, $\rho_i$ is negligible except for four sites 
forming a lattice of charges as close as possible to the Wigner crystal 
triangular network in the imposed square lattice. The degeneracy of the 
crystal is removed by the disorder, the array being pinned in $4$ sites 
of favourable energies.  
For the same cluster, we have calculated 
$C(r)=N^{-1} \sum_i \rho_i \rho_{i+r}$ and used the parameter 
$\gamma=\max_{\,r} C(r) - \min_{\,r} C(r)$ to  
characterize the melting of the crystal 
($\gamma=1$ for a crystal and $0$ for a liquid). 
Calculated for the ground state and the first excited state, $\gamma$ 
allows us to identify 
the second crossing with the melting of the crystal. Moreover, one can 
see that the crystal becomes unstable in the intermediate phase, while 
the ground state is related to the first excitation of the crystal.  
Around the crossings, the longitudinal 
current $I_l$ and the participation ratio $\xi_s =N^2( \sum_i \rho_i^2)^{-1}$ 
of the ground state (i.e. the number of sites that it occupies) are 
enhanced. The general picture reminds us of   
strongly disordered chains~\cite{sjwp} where 
level crossings are associated to a charge reorganization 
of the ground state and comes with an enhancement of the persistent 
current. 
For $k_fl \geq 1$ ($W=5$),  
the previous level crossings are now almost suppressed by a stronger 
level repulsion and charge crystallization occurs more continously.  
There is instead a broad enhancement of $I_l$ which, in contrast to 
the regime $k_F l<1$, is not accompanied by a corresponding increase 
in the number of occupied sites $\xi_s$.  
One can also notice that, increasing $r_s$, $I_t$ is suppressed before $I_l$.
This can be understood looking at the local currents (see Fig.~\ref{fig2}):
the interaction causes the transition from a disordered array of
currents making closed loops to a plastic flow of correlated currents 
in the longitudinal direction~\cite{avishai}. 
%
%
%
%

\begin{figure} 
\centerline{\epsfxsize=16.0cm\epsfysize=20.0cm\epsffile{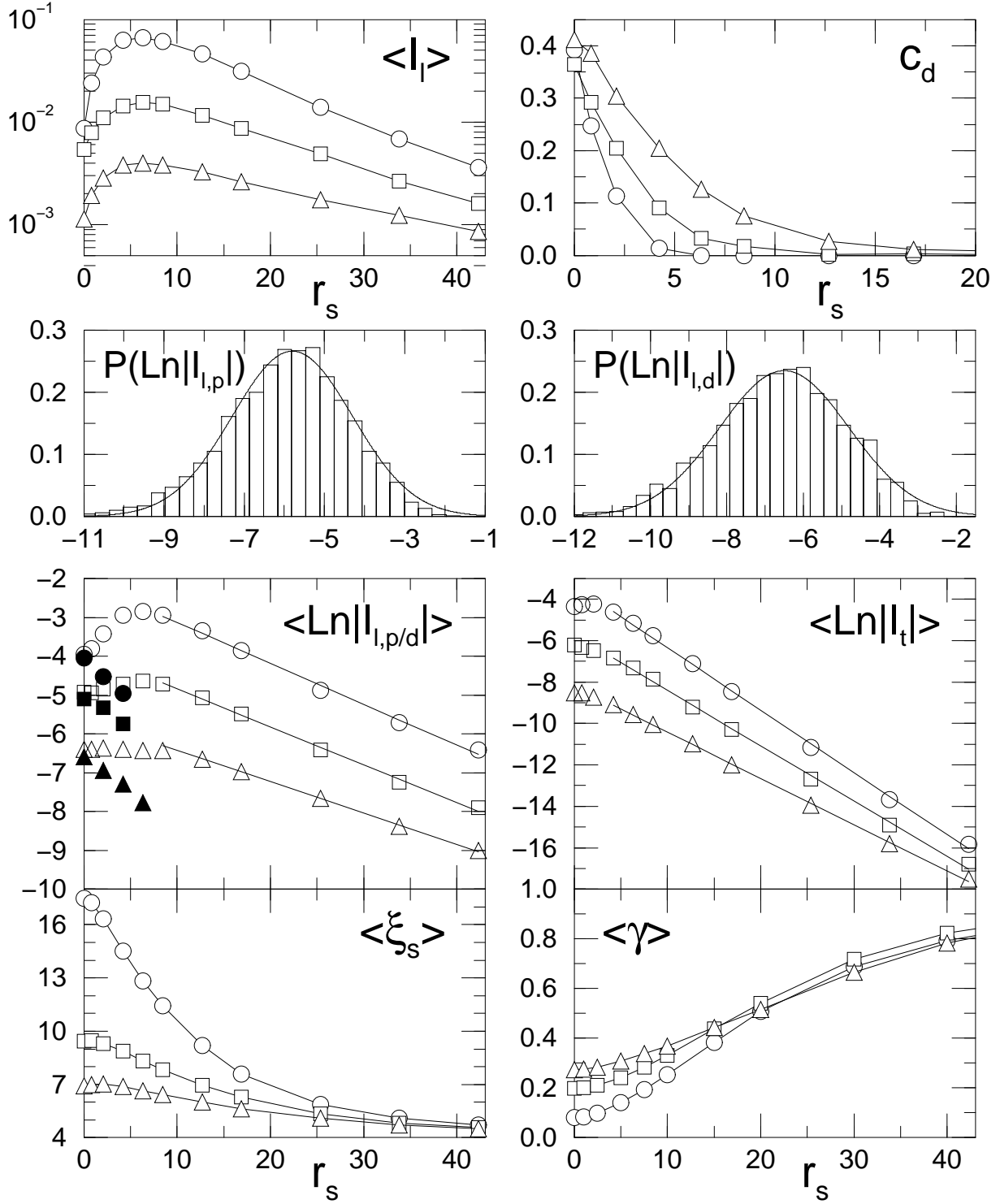}}
\caption{Statistical study of an ensemble of $10^3$ clusters for 
$W= 5$ (circles), $10$ (squares) and $15$ (triangles) as a 
function of $r_s$. 
Mean value of the longitudinal current $<I_l>$.   
Fraction $C_d$ of diamagnetic samples.  
Distribution of the logarithms 
of the paramagnetic currents $P({\rm Ln}|I_{l,p}|)$ 
and of the diamagnetic currents $P({\rm Ln}|I_{l,d}|)$ 
at $r_s=1.7$ and $W=15$ (for an ensemble of $10^4$ clusters), 
fitted by log-normal functions of 
mean $\mu=-5.8$ and variance $\sigma^2=2.3$ in the 
paramagnetic case, $\mu=-6.5$ and $\sigma^2=2.9$ in the 
diamagnetic case. 
Log-averages of the longitudinal paramagnetic (empty symbols) and 
diamagnetic (filled symbols) currents. 
The straight lines are exponential fits giving 
$r_s^W=9.5$ ($W=5$),
$r_s^W=10.3$ ($W=10$),
$r_s^W=12.4$ ($W=15$). 
Log averages of the transverse currents; the exponential fits give
$r_s^F=3.3$ ($W=5$),
$r_s^F=3.7$ ($W=10$),
$r_s^F=4.5$ ($W=15$). 
Mean number of occupied sites $<\xi_s>$. 
Mean crystallization parameter $<\gamma>$. 
\label{fig3}}
\end{figure}

\begin{figure} 
\centerline{\epsfxsize=8.5cm\epsfysize=6.8cm\epsffile{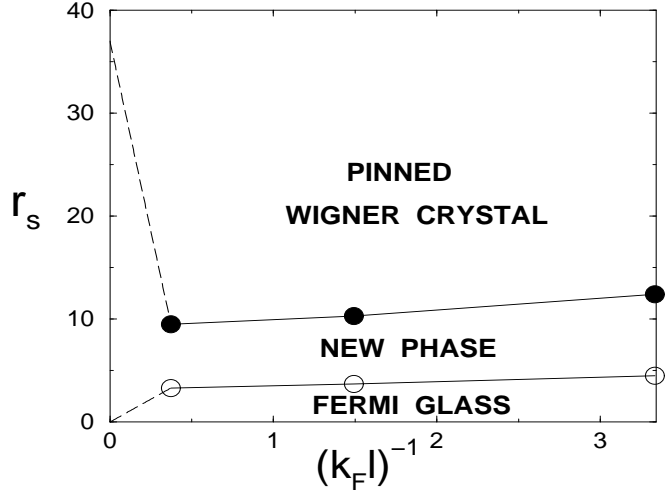}}
\caption{Sketch of the phase diagram for $2d$ spinless fermions in a 
random potential, with $r_s^W$ (filled circle) and $r_s^F$ (empty circle) 
values obtained from Fig.~\ref{fig3}.
\label{fig4}}
\end{figure}

We summarize in Fig.~\ref{fig3} results from a statistical study of 
an ensemble of $10^3$ clusters for $W=5,10,15$.  
One can see an interaction-induced enhancement of the averaged 
persistent current $<I_l>$ 
by almost one order of magnitude when $r_s \approx 7$ and $W=5$. 
This effect can be partially explained by the suppression of the 
fraction of diamagnetic clusters $c_d$.  
The paramagnetic $I_{l,p}$ and 
diamagnetic $I_{l,d}$ longitudinal currents, and $|I_t|$ have 
acceptable log-normal 
distributions for all values of $r_s$ when $W \geq 5$. The stronger is 
the disorder, the better is the log-normal shape of the distributions 
(see Fig.~\ref{fig3} for $W=15$).  
The averages of the logarithms of $I_{l,p}$, $-I_{l,d}$ and $|I_t|$ give 
the typical values shown in Fig.~\ref{fig3}. 
They exponentially decay as $|I_t|\propto \exp(-r_s/r_s^{F})$ 
and $I_l \propto \exp(-r_s/r_s^{W})$ 
when $r_s$ is large enough. 
The values of $r_s^F$ and $r_s^W$ 
extracted from the exponential fits (straight lines of Fig.~\ref{fig3}) 
are given in Fig.~\ref{fig4}, where a sketch of the phase diagram 
is proposed.  
%
%

\section{Conclusions}

A simple model of Coulomb interacting spinless 
fermions in a random potential allows us  
to identify a new quantum phase, clearly separated from 
the Fermi glass and the Wigner crystal. 
This phase is characterized by a plastic flow of currents and a 
response to an Aharonov-Bohm flux with a sign independent of the 
microscopic disorder. 
For $k_Fl \geq 1$, $I_l$ displays a strong 
enhancement for intermediate $r_s$ values, which could be the signature 
of a new metal in the thermodynamic limit.  
For $k_F l<1$, $\xi_s$ and $I_l$ convey similar 
information while the increase of $I_l$ is accompanied by a 
decrease of $\xi_s$ and an increase of $\gamma$  
when $k_Fl > 1$ (see Fig.~\ref{fig3}). This suggests that, in the latter
case, transport in the new phase results from a collective motion 
of charges and not from a delocalization of individual charges.

 This work is partially supported by a TMR network of the EU.

\end{document}